# No-Go Theorems Face Background-based Theories for Quantum Mechanics.

*Louis Vervoort, 02.10.2015*

*Institut National de Recherche Scientifique (INRS),
and Minkowski Institute, Montreal, Canada*
louis.vervoort@emt.inrs.ca

**Abstract**. Recent experiments have shown that certain fluid-mechanical systems, namely oil droplets bouncing on oil films, can mimic a wide range of quantum phenomena, including double-slit interference, quantization of angular momentum and Zeeman splitting. Here I investigate what can be learned from these systems concerning no-go theorems as those of Bell and Kochen-Specker. In particular, a model for the Bell experiment is proposed that includes variables describing a 'background' field or medium. This field mimics the surface wave that accompanies the droplets in the fluid-mechanical experiments. It appears that quite generally such a model can violate the Bell inequality and reproduce the quantum statistics, even if it is based on local dynamics only. The reason is that measurement independence is not valid in such models. This opens the door for local 'background-based' theories, describing the interaction of particles and analyzers with a background field, to complete quantum mechanics. Experiments to test these ideas are also proposed.

*1. Introduction*.

Recently experiments by Couder, Fort and collaborators have demonstrated that fluid-dynamical systems can strikingly mimic quantum behavior [1-4]. Droplets bouncing on a fluid film can exhibit double-slit interference, quantization of angular momentum, the analogue of tunneling and Zeeman splitting, etc. Such droplets, guided by a surface wave, have therefore been called the first macroscopic realizations of a "particle + pilot wave" [1-4]. This impressive series of analogies between fluid-dynamical and quantum mechanical systems naturally sparks the question whether quantum mechanics could, at least in principle, be described by a classical theory – say (the formal equivalent of) fluid dynamics. This question was already posed by Madelung in 1927 [5], who showed that the 1-particle Schrödinger equation can be interpreted as a set of two fluid-mechanical equations. The fluid-mechanical interpretation of quantum mechanics was subsequently elaborated by authors as Wilhelm [21] and Bohm and Vigier [22]; it recently regained interest for the modeling of certain experiments (cf. e.g. [27-28]). However, efforts to derive quantum mechanics from deterministic theories are usually considered futile, in view of various no-go theorems as those of Bell and Kochen-Specker [6-8]. As is well known, these theorems strictly prohibit that local theories describe or complete quantum mechanics. As any bona fide physical theory, fluid dynamics is 'local' in the physical sense that matters here, namely Bell's sense [6]: it (obviously) involves only subluminal, Lorentz-invariant interactions between the fluid elements.

The aim of the present article is investigate what we can learn from the experiments of Couder et al. [1-4] in the context of these no-go theorems. It will be shown that both theorems cannot be applied to 'background-based' or 'fluid-dynamical' theories in general, leaving open the door for such theories to complete quantum mechanics. The reason is that one of the premises on which the theorems are based, often termed measurement independence (MI), is not generally valid in background-based theories, even if these are manifestly local. That measurement independence is a particularly subtle premise of Bell's theorem has been pointed out before (see [14-15, 19, 25, 29, 31] and refs. therein). Refs. [14-15] give a review of MI-violating models reproducing the quantum correlation of the Bell experiment, *in a mathematical and information-theoretic context*. These models have however no known physical interpretation, which is our focus here. Another way to understand our results is that Couder-type systems as well as our models exhibit large-scale correlations at spacelike distances. It was recently argued by G. 't Hooft [20, 31] and others [19] that in such highly correlated systems Bell's theorem might fail. We analyze here in detail how such failure might come about.

Below analysis entirely rests on a careful study of the results presented in [1-4]. For our purposes, then, we need to draw following lessons from these experiments; more details are given in the Appendix. First, the quantum-like behavior is induced by the interaction of the droplets with a 'pilot wave' that guides their horizontal movement. This pilot wave is a (regularly structured) surface wave that results from an external vertical vibration imposed on the fluid film plus the back-reaction of the droplet on the film surface. Under very stringent experimental conditions the system as a whole exhibits a stable probabilistic pattern (one might say that an essential part of the art of the experimenters consisted in identifying these conditions of probabilistic stabilization). For instance, Fig. 1 in Ref. [3] shows the complex phase diagram of the different types of movement of the droplets, and Fig. 17 [3] shows that in the stable 'walking' regime the wave field under the walking droplet can be approximated by simple Fresnel-Huygens theory; it has therefore a high degree of symmetry. Importantly, this stable regime is characterized by *large-scale correlations between the properties of the subsystems*. For instance, in general the height or velocity of a droplet at a given spacetime point is in this regime strongly correlated with the height or velocity of the fluid film (or of the droplet itself) *also at different spacetime points*. This is nicely and amply demonstrated in [1-4], and is recalled in the Appendix. Of course, such massive correlation, potentially between any two properties / variables, is not really a surprise in a system driven by wave dynamics, presenting structure and symmetry, more generally in a stable probabilistic regime.

In a background-based or fluid-dynamical model for the Bell experiment, the Bell particles and analyzers will interact with a background field (describing a fluid-like medium), just as in the experiments of Couder et al. the droplets interact with a fluid's surface field. We thus ask following question: what is the most general mathematical framework for a Bell experiment in which particles and analyzers interact with a fluid ? It will appear we only need probability theory to describe this interaction; the stochastic models we will present here are straightforward generalizations of Bell's assumptions. Clearly, we could more generally invoke a 'background medium / field' instead of a 'fluid'; all we will suppose in the following is interaction of the Bell particles and analyzers with some background (a dark field, the ether, the quantum vacuum,…). In the following stochastic variables (X) are generally n-tuples (n-vectors); two variables X and Y are correlated if the probability P(X|Y) ≠ P(X) for at least some of the values of X and Y. Such a reciprocal relation of probabilistic dependence between X and Y is indicated in a 'correlation graph' by a line between both variables, as in Figs. 1-4 (a general reference on the use of such graphs in statistics is [9]). Let us start from Bell's model, and gradually generalize it.

## *2. Bell's model (M1)*

The probabilistic dependencies assumed in Bell's local hidden-variable (HV) model are schematized in graph M1 (Fig. 1). The spin measurement on the left side ($\sigma_1 = \pm 1$) is supposed to be determined by the left analyzer at angle 'a', and by some set of HVs λ (symbols '$\sigma_2$' and 'b' refer to the right wing). Thus the probability $P(\sigma_1|a,\lambda)$ is assumed to be defined and different from the unconditional $P(\sigma_1)$ in general, i.e. for at least one value of the variables ($\sigma_1$, λ, a).

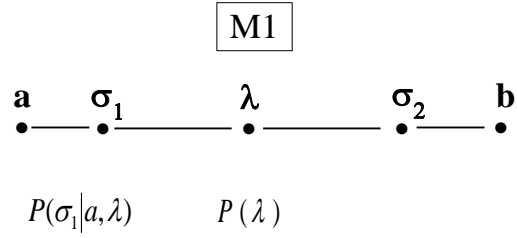

Fig. 1. Correlation graph of Bell's model M1.
The assumed probabilities are indicated (for the left wing).

The HVs λ describe (dynamical) properties of the particle pair, and are characterized by a probability distribution P(λ) taken at a suitably chosen instant; the λ can also be split in left- and right variables [6]. The two essential hypotheses Bell assumes are (1) locality, expressed in stochastic systems by the Clauser-Horne factorability condition [10]:

$$P(\sigma_1,\sigma_2|a,b,\lambda) = P(\sigma_1|a,\lambda) P(\sigma_2|b,\lambda) \quad \text{for all } (\lambda,\sigma_1,\sigma_2), \tag{1}$$

and (2) the condition usually termed measurement independence (MI):

$$P(\lambda|a,b) = P(\lambda|a',b') = P(\lambda) \text{ for all } (\lambda,a,b,a',b') \quad \text{(MI)}. \tag{2}$$

Note that assumptions (1) and (2) are in agreement with the correlation graph[1]. MI (2) follows from the usual 'no-conspiracy' assumption and locality (in advanced experiments the variables (λ, a, b) are mutually spacelike separated) [11-15]. The essential quantity needed to calculate the Bell inequality is the joint probability $P(\sigma_1,\sigma_2|a,b)$, which is in model M1, using (1-2) and assuming discrete values of λ:

$$P^{M1}(\sigma_1,\sigma_2|a,b) = \sum_\lambda P(\sigma_1|a,\lambda)P(\sigma_2|b,\lambda)P(\lambda). \tag{3}$$

Based on (1-3) one proves [7, 10] that in model M1 the Bell inequality (BI) holds:

$$X_{BI}^{M1}(a,a',b,b') = M(a,b)+M(a',b)+M(a,b')-M(a',b') \leq 2 \quad \forall (a,a',b,b'), \tag{4}$$

where the average product $M(x,y) \equiv \langle\sigma_1.\sigma_2\rangle_{x,y} = \sum_{\sigma_1\sigma_2} \sigma_1.\sigma_2 P(\sigma_1,\sigma_2|x,y)$. Quantum mechanics predicts however, in case of the singlet state:

$$P^{QM}(\sigma_1,\sigma_2|a,b) = \frac{1}{4}\left[1-\sigma_1.\sigma_2.\cos(a-b)\right], \tag{5}$$

---

[1] To be precise, by definition correlation graphs represent the correlations that are assumed in the model and that are needed for calculating P(σ₁,σ₂|a,b) (cf. (3)) and thus for verifying the Bell inequality. There may be more correlations in the model. E.g. in Bell's HV model M1 P(σ₁|σ₂) may obviously exist. (In quantum mechanics this probability makes, strictly speaking, no sense; a and b should be defined.)

which strongly violates inequality (4) for certain angles (a,a',b,b').

## *3. Naïve Background Model (M2)*

The first, simplest attempt to model the interaction of a background fluid / field with the Bell particles and analyzers is schematized in the graph of Fig. 2. Whereas λ in M1 describes the particle pair, now $\lambda_1$ and $\lambda_2$ represent *stochastic properties (field intensities,…) of the background medium in the neighborhood of the analyzers*. In model M2 the left spin is determined by the analyzer characteristics a and also by $\lambda_1$, since the particle is supposed to locally interact with the background field – cf. Fig. 2. This simply mimics the droplet-experiments [1-4], in which the droplet and surface field interact and are therefore correlated (cf. Appendix). Furthermore, by analogy with fluids, the properties of the background medium ($\lambda_1$) are supposed to be locally influenced by the characteristics of the analyzer (a) – as when a fluid's surface wave interacts with an obstacle.

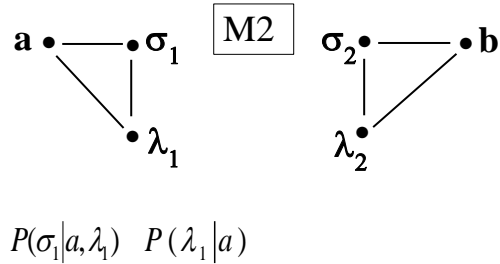

Fig. 2. Correlation graph of model M2.
The assumed probabilities are indicated (for the left wing).

Thus both conditional probabilities $P(\sigma_1|a,\lambda_1)$ and $P(\lambda_1|a)$ are, in general, different from their unconditional counterparts (similarly on the right). As the graph shows there is no further correlation assumed between $\lambda_1$ and $\lambda_2$ so that:

$$P(\lambda_1,\lambda_2|a,b) = P(\lambda_1|a)P(\lambda_2|b), \quad \forall \ (\lambda_1,\lambda_2,a,b), \tag{6}$$

which is compatible with locality, or the spacelike separation between ($\lambda_1$,a) and ($\lambda_2$,b). *A priori* it is worthwhile to investigate model M2, because measurement independence (MI) does not hold in this system: P($\lambda_1$,$\lambda_2$|a,b) ≠ P($\lambda_1$,$\lambda_2$|a',b') ≠ P($\lambda_1$,$\lambda_2$) in general, due to the interactions $\lambda_1$↔a and $\lambda_2$↔b. Since one of the premises of the BI does not hold, the latter is *potentially* violated – but this needs to be calculated (the two conditions (1-2) are jointly sufficient for the BI to hold in the model, but they are not necessary). Instead of Eq. (3) one sees that we have in M2:

$$P^{M2}(\sigma_1,\sigma_2|a,b) = \sum_{\lambda_1,\lambda_2} P(\sigma_1,\sigma_2|\lambda_1,\lambda_2,a,b)P(\lambda_1,\lambda_2|a,b)$$
$$= \sum_{\lambda_1,\lambda_2} P(\sigma_1|a,\lambda_1)P(\sigma_2|b,\lambda_2)P(\lambda_1|a)P(\lambda_2|b). \qquad (7)$$

Even if M2 violates MI, it is still straightforward to prove that it satisfies the BI. Indeed, (7) can be rewritten as $P^{M2}(\sigma_1,\sigma_2|a,b) = P(\sigma_1|a)P(\sigma_2|b)$, which is of the form (3) with a fixed λ. In conclusion:

$$X_{BI}^{M2}(a,a',b,b') \leq 2, \quad \forall\ (a,a',b,b'). \qquad \square \qquad (8)$$

Note that violation of MI in M2 is of course compatible with 'free will'[2]. Remarkably, one only needs to add one type of interaction to M2 to come to the desired result, as we now prove.

*4. Second Background Model (M3)*

The next logical step is to add to M2 a stochastic property that belongs to *both* particles, say represented by the hidden variable(s) ξ. So we keep the HVs ($\lambda_1,\lambda_2$) from M2 describing the background medium, and add the property λ from M1, now denoted ξ, describing the particle pair. Just as in Bell's original model [6], these new properties can be (quasi-)constants of motion or dynamical variables with laws of motion; ξ can then again be thought of as values at some suitable moment. In the most general setting the left [right] spin will depend on (ξ, $\lambda_1$, a) [(ξ, $\lambda_2$, b)]: an almost trivial variant of Bell's assumptions.

---

[2] Sure, violation of MI implies that P(a,b|$\lambda_1,\lambda_2$) ≠ P(a,b) in general, which some might (wrongly) interpret as an impossible 'causal determination' of (a,b) by HVs ($\lambda_1,\lambda_2$) – impossible because (a,b) can be freely or randomly chosen in some experiments. But P(a,b|$\lambda_1,\lambda_2$) is a meaningful measure itself; this is most easily understood for series of experiments in which (a,b) are stochastic variables themselves, i.e. take different values with a given distribution P(a,b) – as happens in modern Bell experiments. As another example, think of P(x|T) with x = half-life of an excited molecular state, T = experimental temperature (suppose that a few discrete values of x and T are sampled). If one performs a large number of experiments measuring x at different T's, P(T|x) ≠ P(T) in general even if one can choose T freely.

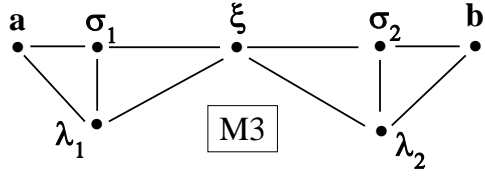

$P(\sigma_1|\xi,\lambda_1,a) \ P(\xi|\lambda_1,\lambda_2) \ P(\lambda_1|a)$

Fig. 3. Correlation graph of the 'fluid' model M3.
The assumed probabilities are indicated (for the left wing).

For all the following results we further only need to suppose that the values of $\lambda_1$ and $\lambda_2$ depend in general on $\xi$ (and hence vice versa), i.e. $P(\lambda_1,\lambda_2|\xi) \neq P(\lambda_1,\lambda_2)$ for at least some values of $(\lambda_1,\lambda_2,\xi)$. Clearly, this is possible in a model in which the particles (characterized by $\xi$) interact with the background fluid $(\lambda_1,\lambda_2)$. E.g., in the Paris experiments the properties of the fluid film, say its shape and surface velocity field, are strongly dependent on the properties of the droplets, say their mass, velocity or trajectory [1-4]. That correlations of the type $P(\lambda_1,\lambda_2|\xi)$ can legitimately be assumed to exist is further illustrated in the Appendix (cf. in particular Eq. (A1)); by Bayes' rule it then follows that correlations $P(\xi|\lambda_1,\lambda_2)$ will also, in general, exist. We end up with graph M3 in Fig. 3, which combines M1 and M2.

Instead of Eqs. (3) and (7) we have in M3:

$$\begin{aligned} P^{M3}(\sigma_1,\sigma_2|a,b) &= \sum_{\lambda_1,\lambda_2} P(\sigma_1,\sigma_2|\lambda_1,\lambda_2,a,b)P(\lambda_1,\lambda_2|a,b) \\ &= \sum_{\lambda_1,\lambda_2}\sum_{\xi} P(\sigma_1,\sigma_2|\xi,\lambda_1,\lambda_2,a,b)P(\xi|\lambda_1,\lambda_2,a,b)P(\lambda_1,\lambda_2|a,b) \\ &= \sum_{\lambda_1,\lambda_2}\sum_{\xi} P(\sigma_1|\xi,\lambda_1,a)P(\sigma_2|\xi,\lambda_2,b)P(\xi|\lambda_1,\lambda_2)P(\lambda_1|a)P(\lambda_2|b). \end{aligned} \quad (9)$$

The essential difference with M2 is the presence of the factor $P(\xi|\lambda_1,\lambda_2)$ in (9). Mathematically, the only constraints on the probabilities in (9) (besides their being $\in [0,1]$) are the following normalization conditions:

$$\sum_{\sigma_1} P(\sigma_1|\xi,\lambda_1,a) = 1, \quad \forall (\xi, \lambda_1, a), \text{ and similarly for } \sigma_2, \quad (10a)$$

$$\sum_{\xi} P(\xi|\lambda_1,\lambda_2) = 1, \quad \forall (\lambda_1, \lambda_2), \quad (10b)$$

$$\sum_{\lambda_1} P(\lambda_1|a) = 1 = \sum_{\lambda_2} P(\lambda_2|b), \quad \forall (a,b). \quad (10c)$$

In (9) we have assumed that the Clauser-Horne locality (1) is satisfied (with $\lambda$ replaced by $(\xi,\lambda_1,\lambda_2)$), in agreement with the graph. However, MI is violated in M3:

$$\frac{P(\lambda_1,\lambda_2,\xi|a,b)}{P(\lambda_1,\lambda_2,\xi|a',b')} = \frac{P(\lambda_1,\lambda_2|a,b).P(\xi|\lambda_1,\lambda_2,a,b)}{P(\lambda_1,\lambda_2|a',b').P(\xi|\lambda_1,\lambda_2,a',b')}$$

$$= \frac{P(\lambda_1|a).P(\lambda_2|b).P(\xi|\lambda_1,\lambda_2)}{P(\lambda_1|a').P(\lambda_2|b').P(\xi|\lambda_1,\lambda_2)} = \frac{P(\lambda_1|a).P(\lambda_2|b)}{P(\lambda_1|a').P(\lambda_2|b')} \neq 1, \quad (11)$$

in general, due to the interactions $\lambda_1 \leftrightarrow a$ and $\lambda_2 \leftrightarrow b$.

Based on (9), it is straightforward to prove that the BI can be maximally violated in model M3. To that end, it suffices to treat the simplest case, in which the HVs $\lambda_1, \lambda_2, \xi$ are scalar properties each taking just two values, which one can choose to be (1,2):

$$\lambda_1, \lambda_2, \xi = 1,2. \quad (12)$$

The average products $M(x,y)$ that compose $X_{BI}$ can be written as follows:

$$M(x,y) = 2\left[P(\sigma_1=+1,\sigma_2=+1|x,y) + P(\sigma_1=-1,\sigma_2=-1|x,y)\right] - 1. \quad (13)$$

Using (9) and (12), one obtains:

$$P(\sigma_1=+1,\sigma_2=+1|a,b) \equiv P(\sigma_1^+,\sigma_2^+|a,b)$$
$$= \sum_{\lambda_1,\lambda_2} P(\sigma_1^+|\xi=1,\lambda_1,a)P(\sigma_2^+|\xi=1,\lambda_2,b)P(\xi=1|\lambda_1,\lambda_2)P(\lambda_1|a)P(\lambda_2|b) +$$
$$+ \sum_{\lambda_1,\lambda_2} P(\sigma_1^+|\xi=2,\lambda_1,a)P(\sigma_2^+|\xi=2,\lambda_2,b)P(\xi=2|\lambda_1,\lambda_2)P(\lambda_1|a)P(\lambda_2|b). \quad (14)$$

To maximize $X_{BI}$, it appears sufficient to judiciously choose the probabilities $P(\xi|\lambda_1,\lambda_2)$, $P(\lambda_1|a)$, $P(\lambda_2|b)$ in (14) equal to 0 or 1, while satisfying (10a-c). Specifically, take $P(\lambda_1=1 | a) = 1$, $P(\lambda_2=1 | b) = 1$, $P(\xi=1 | \lambda_1=1, \lambda_2=1) = 1$ (in short: $P(1|a) = P(1|b) = P(1|1,1) = 1$). Then, using the normalization (10b-c), only one term in (14) survives:

$$P(\sigma_1=+1,\sigma_2=+1|a,b) = P(\sigma_1^+|\xi=1,\lambda_1=1,a)P(\sigma_2^+|\xi=1,\lambda_2=1,b) \equiv \alpha_1 \alpha_2, \quad (15)$$

defining the parameters $\alpha_i$ ($\in [0,1]\ \forall\ i$) which we can freely choose. Hence the second term in (13) becomes:

$$P(\sigma_1=-1,\sigma_2=-1|a,b) = P(\sigma_1^-|\xi=1,\lambda_1=1,a)P(\sigma_2^-|\xi=1,\lambda_2=1,b) = (1-\alpha_1)(1-\alpha_2). \quad (16)$$

Therefore $M(a,b)$ in (13) is:

$$M(a,b) = 2\,[\,\alpha_1.\alpha_2 + (1-\alpha_1)(1-\alpha_2)\,] - 1, \quad (17)$$

which is maximized to 1 by choosing e.g. $\alpha_1 = \alpha_2 = 1$. It is easy to see that for each of the 4 couples of angles (x,y) intervening in $X_{BI}$ (cf. (4)) $M(x,y)$ can be written as in (17) with two new degrees of freedom ($\alpha_1(x,y), \alpha_2(x,y)$) for each couple (x,y). For instance, if we choose for (a,b'): $P(2|b') = 1 = P(1|a)$ (the latter probability already being fixed above) and $P(2|1,2) = 1$, we find as in (15) that:

$$P(\sigma_1 = +1, \sigma_2 = +1 | a, b') = P(\sigma_1^+ | \xi = 2, \lambda_1 = 1, a) P(\sigma_2^+ | \xi = 2, \lambda_2 = 2, b') \equiv \alpha_3 \alpha_4. \qquad (18)$$

In sum, to calculate M(x,y) we can choose $P(\xi|\lambda_1,\lambda_2)$, $P(\lambda_1|x)$, $P(\lambda_2|y) \in \{0,1\}$ in such a way that the vector $(\xi, \lambda_1, x)$ is different for the 4 couples (x,y) (even if x is the same for 2 couples); idem for $(\xi, \lambda_2, y)$. Therefore also $P(\sigma_1= +1|\xi, \lambda_1, x)$ and $P(\sigma_2= +1|\xi, \lambda_2, y)$ can be chosen differently for the 4 couples (x,y); we have $2^3 = 8$ DOF $(\alpha_1,…\alpha_8)$. Finally, using (4) and (13) these choices lead to:

$$X_{BI} = 2 \left[ \alpha_1\alpha_2 + (1-\alpha_1)(1-\alpha_2) + \alpha_3\alpha_4 + (1-\alpha_3)(1-\alpha_4) + \right.$$
$$\left. + \alpha_5\alpha_6 + (1-\alpha_5)(1-\alpha_6) - \alpha_7\alpha_8 - (1-\alpha_7)(1-\alpha_8) \right] - 2. \qquad (19)$$

This implies $X_{BI} = 4$ e.g. for the choice: all $\alpha_i = 1$ except $\alpha_8 = 0$ (as well as for 15 other choices). In conclusion, there are background-based models conceivable of type M3 (i. e. satisfying (9-10)) for which:

$$X_{BI}^{M3}(a,a',b,b') > 2, \quad \text{for some } (a,a',b,b'). \qquad \square \qquad (20)$$

Some such models, in particular the 'dichotomic' background model satisfying (12), can maximally violate the BI up to $X_{BI} = 4$.

The next interesting question is whether it is possible that M3-models not only violate the BI, but that $P^{M3}(\sigma_1,\sigma_2|a,b)$ in (9) coincides with $P^{QM}(\sigma_1,\sigma_2|a,b)$ (Eq. (5)). This appears indeed possible, for some choices of the probabilities in (9), as we will prove using a result obtained by Hall [14]. Let us call the Hall model M4; its graph is given in Fig. 4. In [14] it is proven that a particular instance of this graph can reproduce the quantum correlation through violation of MI only. (Other models may violate locality (2); for a review of mathematical and information-theoretic models reproducing (5) see [14-15].)

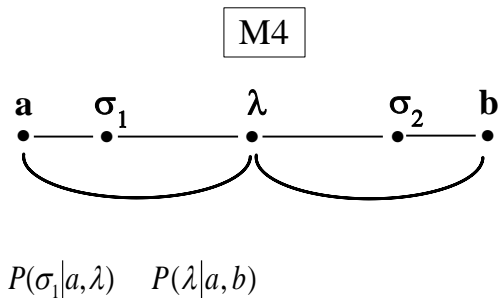

Fig. 4. Correlation graph of Hall's model M4.
The assumed probabilities are indicated (for the left wing).

As one sees on the graph, MI-violation in this model comes about through correlation of the analyzer variables a and b with the variables λ describing the particle pair. On the standard interpretation, such a correlation is understood as a non-local effect or a cosmic conspiracy (since the events of setting 'a' and 'b' are spacelike separated from the emission event determining the λ-values, at least in the most advanced experiments, e.g. [11-13]). In detail, the model of Ref. [14] assumes that (a,b) and λ are continuous 3-vectors on the unit sphere, and that:

$$P(\sigma_1|\lambda,a) = \delta_{\sigma_1,A(\lambda,a)} \text{ with } A(\lambda,a) = \text{sgn}(a \cdot \lambda) \qquad (21a)$$

$$P(\sigma_2|\lambda,b) = \delta_{\sigma_2,B(\lambda,b)} \text{ with } B(\lambda,b) = -\text{sgn}(b \cdot \lambda) \qquad (21b)$$

$$\rho(\lambda|a,b) = \frac{1+a \cdot b}{8(\pi-\varphi_{a,b})} \text{ for } \text{sgn}(a \cdot \lambda) = \text{sgn}(b \cdot \lambda)$$

$$= \frac{1-a \cdot b}{8\varphi_{a,b}} \text{ for } \text{sgn}(a \cdot \lambda) \neq \text{sgn}(b \cdot \lambda). \qquad (21c)$$

Here $\delta_{\bullet,\bullet}$ is the Kronecker-δ ; the functions A and B ∈ {−1,+1}. Further $\varphi_{a,b}$ ∈ [0, π] is the angle between the vectors a and b; the density ρ in (21c) is defined as zero when the denominators vanish. Finally, the Clauser-Horne locality condition (1) is assumed, which is compatible with the choices (21a-b) and with the graph of Fig. 4. Note that (21c) implies that MI in (2) is violated; (21a-c) lead to the quantum result (5).

We can formally reduce the correlation function of fluid model M3 to the one of model M4 by following procedure. Assume that $\lambda_1$, $\lambda_2$ and $\xi$ are also continuous 3-vectors on the unit sphere, and make following normalized choices for the probabilities in (9):

$$P(\lambda_1|a) = \delta(\lambda_1 - a), \qquad P(\lambda_2|b) = \delta(\lambda_2 - b) \qquad (22a)$$

$$P(\sigma_1|\xi,\lambda_1,a) = \delta_{\sigma_1,A(\xi,\lambda_1,a)} \text{ with } A(\xi,\lambda_1,a) = \text{sgn}(a \cdot \lambda_1)\text{sgn}(a \cdot \xi) \qquad (22b)$$

$$P(\sigma_2|\xi,\lambda_2,b) = \delta_{\sigma_2,B(\xi,\lambda_2,b)} \text{ with } B(\xi,\lambda_2,b) = -\text{sgn}(b \cdot \lambda_2)\text{sgn}(b \cdot \xi) \qquad (22c)$$

$$\rho(\xi|\lambda_1,\lambda_2) = \frac{1+\lambda_1 \cdot \lambda_2}{8(\pi-\varphi_{\lambda_1\lambda_2})} \text{ for } \text{sgn}(\xi \cdot \lambda_1) = \text{sgn}(\xi \cdot \lambda_2)$$

$$= \frac{1-\lambda_1 \cdot \lambda_2}{8\varphi_{\lambda_1\lambda_2}} \text{ for } \text{sgn}(\xi \cdot \lambda_1) \neq \text{sgn}(\xi \cdot \lambda_2). \qquad (22d)$$

Here $\varphi_{\lambda_1\lambda_2}$ ∈ [0, π] is again the angle between the vectors $\lambda_1$ and $\lambda_2$; the density ρ in (22d) is defined as zero when the denominators vanish. By projecting $\lambda_1$ on a and $\lambda_2$ on b via (22a), one projects (22b-d) onto (21a-c). Explicitly, with (22a-c), by integrating over $\lambda_1$ and $\lambda_2$ in (9) one finds:

$$P^{M3}(\sigma_1,\sigma_2|a,b) = \int \delta_{\sigma_1,A(\xi,a,a)}\delta_{\sigma_2,B(\xi,b,b)}\rho(\xi|a,b)d\xi , \qquad (23)$$

which is the expression of which Hall has proven that it leads to the desired quantum result (5) [14]. □

We thus see that a simple 3-variable model as M3, involving background variables $\lambda_1$ and $\lambda_2$, has enough degrees of freedom to reproduce the Bell-correlation. Now, even if for certain choices of probabilities the correlation function of M3 is identical to that of M4, the physics behind both models is entirely different, as is also apparent from their graphs. While the graph of Hall model M4 can come about through direct delocalized interaction between (a,b) and $\lambda$, background model M3 does nowhere rely on delocalized or nonlocal interactions; all interactions can be local, as in a fluid. Therefore one expects that the graph of M3, contrary to that of M4, can survive under conditions of Einstein locality, i.e. in dynamic experiments [11-13]. Indeed, in such advanced experiments any correlation in the graph of M4 between $\lambda$ and a and between $\lambda$ and b is destroyed (unless one assumes nonlocality or conspiracy) since $\lambda$, a and b are mutually spacelike separated. Spacelike separation is achieved by randomly switching the analyzer directions at high enough frequencies, prohibiting the exchange of signals and ensuring that MI is satisfied. However, this experimental procedure cannot work for M3: one cannot separate 'a' from its nearby environment ($\lambda_1$), and similarly on the right side, at least not at the frequencies used to separate the left and right wings in the existing experiments[3] [11-13]. Thus violation of MI in (11) is not excluded. If that is correct, *the graph of M3 can also exist in dynamic experiments, without assuming any nonlocality or conspiracy*. Recall that also correlations of the type $\rho(\xi|\lambda_1,\lambda_2)$ can exist at the time of measurement ($t_m$), simply because the particles (described by $\xi$) can interact with the background medium ($\lambda_1$, $\lambda_2$) in the neighborhood of the analyzers; thus $\xi(t_m)$ partly determines $\lambda_1(t_m)$ and $\lambda_2(t_m)$ (in the probabilistic sense). Indeed, let us repeat that for our purposes the essential lesson from the experiments of Couder et al. is this: in fluid-dynamical systems characterized by a large-scale probabilistic regularity, properties of a subsystem at a given spacetime point remain strongly correlated with properties of the same *and* of other subsystem(s) *even at different* spacetime points – in principle the spacetime regions can be indefinitely separated. The examples provided by the experiments of Refs. [1-4] are numerous and convincing, and see the Appendix. Hence the factor $P(\xi|\lambda_1,\lambda_2)$ in (9), expressing large-scale correlations, remains meaningful: the property ξ keeps on describing both particles throughout their flight and partly determines ($\lambda_1$, $\lambda_2$), even if $\lambda_1$ ($\lambda_2$) is also influenced by a now varying angle a (b)[4].

---

[3] For instance, in Ref. [12] the switching frequency f is maximum 30 MHz, amply sufficient to causally separate the left and right wings which were 400 m apart. This frequency amounts to a causal range $\Delta x = c/f = 10$ m. Note that within model M3, all $\lambda_1$ and $\lambda_2$ within this range $\Delta x$ are, at the time of measurement, in causal contact with their respective analyzer variable.

## 5. Interpretation and Suggested Experiments

If model M3 is a faithful be it simplified abstraction of what happens in the droplet-system of Couder, Fort et al., then it explains how in strongly correlated systems violation of MI (and the BI) can happen in a physically totally harmless manner. *MI-violation simply arises through the interaction of analyzers (a, b) and background ($\lambda_1$, $\lambda_2$), as in Eq. (11).* Along these lines, G. 't Hooft has recently argued that measurement dependence (termed 'conspiracy' in [31]) may be due to bona-fide correlations between spacelike separated events [31]. The author has provided several quantum models that can be mapped to classical systems but typically exhibit such large correlations at spacelike distances [31]. Our result can thus be seen to be in qualitative agreement with 't Hooft's conclusion; certain fluid-mechanical systems as Couder's can exhibit such large correlations at spacelike separation (see e.g. Fig. 5 in the Appendix).

Violation of MI is an example of what we termed 'supercorrelation' in Ref. [19], i.e. correlations that are stronger than allowed by Eqs. (1-2). In [19] we showed that in spin-lattices such supercorrelation exists, in particular violation of MI, leading to violation of the BI for certain lattices. For our present purpose it is interesting to note that spin-lattices appear to be closely related to the background-based models *à la* M3 studied here. Violation of MI and the BI come about in such lattices through interaction of the analyzer-spins and test-spins with a stochastic background medium (namely the remaining spins in the lattice) [19]. These systems thus give a further physical basis to our model. Another particularly relevant physical system is given in Ref. [18], which argues that vortices in fluids can be precisely correlated as electrons in a singlet state. Although [18] does not provide a hidden-variable model, the findings of this work and ours corroborate each other (see [16-17] for other recent work inspired by the droplet experiments [1-4]). Finally, and needless to say, we believe it is now highly desirable to revisit, and investigate the precise link with, the pilot-wave theory of de Broglie and Bohm (in particular in its hydrodynamic formulation); and also sub-quantum theories *à la* Nelson, which invoke a Brownian background medium [23-24, 26].

---

[4] An interesting question is whether the factorability condition $P(\lambda_1, \lambda_2|\xi) = P(\lambda_1|\xi) \, P(\lambda_2|\xi)$ or equivalently $P(\lambda_2|\xi, \lambda_1) = P(\lambda_2|\xi)$ must be violated in model M3 (as in Eq. (22d)) in order to violate the BI. Strictly speaking this is not so, because full asymmetric decoupling of $\lambda_2$ (or $\lambda_1$), i.e. $P(\lambda_2|\xi, \lambda_1) = P(\lambda_2|\xi) = P(\lambda_2)$ can be shown to lead to violation of the BI, using a result obtained in [30]. However, the question remains for the physically realistic case where the background symmetrically interacts with the particles. At any rate, such 'non-factorizing' correlations do exist in nature, as the reader will easily infer from the examples given in the Appendix.

Before turning to ways to experimentally test these ideas, a few other remarks are in place. First, note that also the Kochen-Specker theorem assumes measurement independence, as is well explained in [15], p. 11. In discussions of the latter theorem MI is usually termed 'non-contextuality', the property that physical quantities ('elements of reality') are *in*dependent on the measurement settings. Since background-based or fluid-dynamical models of type M3 (i.e. satisfying Eqs. (9) and (10a-c)) violate this premise, also the Kochen-Specker theorem cannot rule out such models. Second, one will observe that a variety of other types of graphs than M3 are conceivable; we have here only investigated the simplest variant. A classification of such graphs may be interesting but goes beyond the aim of this article. Let us here just note that the more correlations one includes, the more resources one has to reproduce the quantum result [14-15, 19].

Although speculative, it is tempting to conceive experiments to test the ideas presented here, in fluid-mechanical but possibly also in quantum systems. First of all, it may be feasible to devise a Bell-type experiment on the droplet systems, using pairs of droplets. One would have to measure a dichotomic property that depends on one of the many control parameters, which will play the role of analyzer variable in the Bell inequality. The first experimental challenge would be, it seems, to prepare pairs of droplets in a sufficiently correlated state while both particles move in roughly opposite directions (note that such a correlated state has been created for two co-rotating droplets [4]). If feasible our results indicate that such an experiment could violate Bell-type inequalities via the predicted violation of MI. One would need to find the right experimental conditions and fine-tune the existing correlations using all control parameters. Nevertheless, in view of the massive correlation that exists in these systems one is tempted to conjecture that under optimized conditions the BI will be violated in this classical system.

An avenue to test our model in the quantum realm might be to modify existing Bell experiments that use high-frequency switching of the analyzers [11-13]. Suppose that the analyzers indeed interact with a background field / medium; then by sufficiently increasing the switching frequency, beyond the presently used ones, one expects that at some point the analyzers will be decoupled *also from their nearby environment*. In that case MI is again satisfied (cf. (11)) and model M3 breaks down. Thus, while within quantum mechanics $X_{BI}$ in (4) is entirely independent of the switching frequency and equal to $2\sqrt{2}$ for the optimum angles, fluid-dynamical and background-based models seem to predict a different behavior, namely a (continuous) evolution of $X_{BI}$ asymptotically reaching $X_{BI} = 2$ at high enough frequencies. Even if it seems difficult to predict at which frequencies such decoupling occurs, it is not excluded that these frequencies are within experimental reach[5]. As another test, one could maybe use effectively *rotating* analyzers, as Bell originally proposed [6]. One expects that from a critical angular velocity range on a background medium ($\lambda$) will at most experience a smeared-out or averaged influence from the whole angular range (including a, a', b, b') through which the analyzers scan. In other words $P(\lambda)$ in (2) (or $P(\lambda_1,\lambda_2)$) may depend on the whole range of angles but not just on (a,b); this is equivalent to MI and corresponds to Bell's argument for using rotating analyzers.

## *6. Conclusion*

We have presented a description of a Bell experiment in which the two particles and analyzers locally interact with a background (a field, a fluid-like medium). Drawing on the essential lessons to be learned from recent fluid-dynamical experiments [1-4], we showed that such models (M3 above) can violate the Bell inequality and reproduce the quantum correlation of the Bell experiment. The 'resource' appears to be violation of measurement independence, a particularly subtle premise of the Bell inequality. We argued that MI-violating mechanisms are compatible with locality and free will in our model; superdeterminism does not need to be invoked. Of course, one may well say that such fluid / background models invoke a (harmless) form of 'delocalized extendedness' as fluids and fields normally do. But such models do not exhibit the pathological nonlocality that Einstein and Bell sought to exclude. All interactions in a fluid, and in model M3, are local in the physically important sense. Besides the background, the second essential assumption of model M3 is the existence of long-range correlations between the left and right parts of the experiment (condensed in the factor $P(\xi|\lambda_1,\lambda_2)$ in (9)). Such correlations are convincingly shown to exist in the Paris droplet experiments [1-4], as recalled in the Appendix. The correlation P(ξ|λ$_1$,λ$_2$), or equivalently P(λ$_1$,λ$_2$|ξ), can also be understood as expressing that the global properties of the pilot-wave (the background field) are determined by the particle pair's properties. Our main conclusions are in qualitative agreement with recent classical models proposed by G. 't Hooft, which exhibit such correlations at spacelike separations and reproduce certain quantum features [31].

---

[5] As an example, if one assumes that the coupling breaks down when the causal range $\Delta x$ becomes of the order of the typical length of the polarizers, say 10 cm, one finds a decoupling frequency of the order of a few GHz, which may technically be in reach (Gregor Weihs, private communication).

Another angle from which these results can be understood is the following. In model M3 the hidden variables (ξ,λ$_1$,λ$_2$) not only describe the particle pair, but also a background field or medium. This broadened meaning of the 'hidden variables' doubtlessly goes beyond Bell's initial model [6], in which he explicitly attached the additional variables to the particle pair. On the other hand, in a more recent publication [7] Bell himself opened the door to a much wider interpretation of what hidden variables might be (cf. [7] p. 52, p. 54-56 and the example discussed there). It is clear from these passages that in this broader framework the hidden variables do not need to pertain to the particle pair alone. Nevertheless the idea that the hidden variables are only particle-related is still overwhelmingly present in the Bell-literature.

On this broader meaning, the statement that 'local hidden-variable theories are impossible', which has almost reached the status of an axiom in modern physics, appears to be untenable. Some authors are already exploring new avenues for concrete theories completing quantum mechanics [20].

*Acknowledgements*. I would like to thank, for instructive discussions, Yves Gingras, Vesselin Petkov, Jean-Pierre Blanchet, Chérif Hamzaoui, Gilles Brassard, David Poulin, Robert Brady, Ross Anderson and Richard MacKenzie.

**Appendix. Note on the correlations existing in the fluid-mechanical system of Refs. [1-4]**

As shown in detail in Refs [1-4], under specific experimental conditions oil droplets can be created that 'walk' over a vibrating oil film. In these conditions the droplets bounce rapidly on the film, thus creating a surface wave that propels them horizontally over the film. A first observation to make is that droplets will only walk if several parameters lie in tight intervals. There are about 10 such control parameters, e.g. the viscosity of the oil droplet and oil film, the mass of the droplet, the geometrical dimensions of the bath, the height of the oil film etc.. Outside these ranges there is no walking: the system is chaotic and no droplets will form or they will be rapidly captured by the film. Second, Couder, Fort et al. have shown that the stable walking regime is *probabilistic* (see e.g. the histograms of the droplet's deviation angle after double-slit interference in Fig. 3, Ref. [2]). In other words the trajectories of two 'identical' droplets starting with 'identical' initial conditions will at best only roughly coincide: there are statistical fluctuations; but probabilistic patterns stabilize and can be repeated and measured.

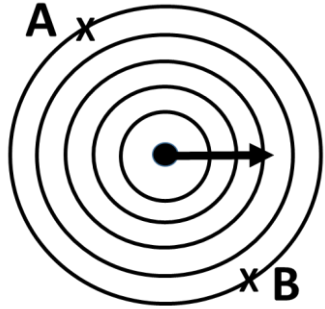

Fig. 5. To first approximation, the wave field under the walking droplet is circular; the figure shows the anti-nodal lines (cf. e.g. Fig. 17 in Ref. [3]).

Third, it is shown in detail by Couder et al. that the droplets in the experiments are guided by symmetric waves, which are in first approximation circular (cf. Fig. 5). A better approximation is given by Fresnel – Huygens theory, and consists of the superposition of the circular waves created by the droplet at each impact on its trajectory (see e.g. Fig. 17 in [3]).

The essential feature of the droplet-experiments we will use is that in the stable regime, there is a strong correlation potentially between *all* system variables. (Note that one could even consider a geometric parameter as the width of the bath as a stochastic parameter: such a variable could parametrize a series of experiments in which this width has different values. The wave properties are also strongly dependent on these geometric parameters.) For instance, in the system sketched in Fig. 5 the height of the wave at A will be strongly correlated with the height at B; we have e.g. $P(h_A | h_B) \approx 1$ if $h_A \approx h_B$; in general $P(h_A | h_B) \neq P(h_A)$ in a series of probabilistic experiments ($h_X$ is the event that the surface height at point X takes on a certain value). Since the droplet closely follows the surface wave (it hits the latter periodically at the same point of the wave front [1-2]), the properties of the droplet will also be strongly correlated with the properties of the surface wave on the fluid film. For instance, the position of a droplet at given time will be correlated with the height of the surface wave also at a different space-time point. More generally, since the precise shape of the wave front is strongly dependent on the droplet properties, e.g. its mass, velocity, etc. [1-4], any surface wave property (at a given space-time point) can be correlated with potentially any of the droplet properties (also at different space-time points).

Finally, under precise conditions two droplets can be strongly correlated: in Ref. [4] it is shown that droplets can bounce in phase or anti-phase, while being trapped in the wave-field they generate (in this case the z-positions of the droplets are *perfectly* correlated). What would happen if two (identical) droplets could be created in the center of a symmetric bath while moving in opposite directions ? Then one expects – or it is certainly a logical possibility – that under certain conditions a highly symmetric surface field will form and again correlate the movements of both droplets (the simplest assumption is that their movements will be perfectly symmetric; but in any case there will be stochastic deviations). The shape and properties of the 2-particle wave field will again depend on the droplet properties, e.g. the mass m of both identical droplets. Therefore, in general we will have correlations of the type

$$P(\lambda_1, \lambda_2 | m) \neq P(\lambda_1, \lambda_2), \tag{A1}$$

where $\lambda_1$ and $\lambda_2$ are properties of the wave field, for instance the height ($\lambda_1$) of the surface wave at some reference point on the left side, and $\lambda_2$ the same property on the right side. Clearly, for different droplet masses these field characteristics will in general be different [1-4].

All these features are a consequence of the probabilistic nature of the system as a whole, and the structured, high-symmetry wave that guides the droplets. When we propose in the main text a model for the interaction of Bell-particles with a background, we only extrapolate correlation properties that exist in the droplet-experiments.

**References**.